# Transition to turbulence and streamwise inhomogeneity of vortex tangle in thermal counterflow

E. Varga · S. Babuin · V. S. L'vov · A. Pomyalov · L. Skrbek

**Abstract** We report preliminary results of the complementary experimental and numerical studies on spatiotemporal tangle development and streamwise vortex line density (VLD) distribution in counterflowing $^4$He. The experiment is set up in a long square channel with VLD and local temperature measured in three streamwise locations. In the steady state we observe nearly streamwise-homogeneous VLD. Experimental second sound data as well as numerical data (vortex filament method in a long planar channel starting with seeding vortices localized in multiple locations) show that the initial build up pattern of VLD displays complex features depending on the position in the channel, but the some tangle properties appear uniform along its length.

## 1 Introduction

Various aspects of quantum turbulence (QT) generated in thermal counterflow of superfluid $^4$He has been the subject of numerous studies since the pioneering experiments and their phenomenological description of Vinen [1]. Along with his studies of steady-state counterflow, he was also interested in transient effects such as temporal development of temperature difference $\Delta T$ and vortex line density (VLD), based on experiments performed with two

E. Varga · L. Skrbek
Faculty of Mathematics and Physics, Charles University, 121 16 Prague, Czech Republic
E-mail: varga.emil@gmail.com

S. Babuin
Institute of Physics ASCR, Na Slovance 2, 180 00 Prague, Czech Republic

V.S. L'vov · A. Pomyalov
Department of Chemical Physics, Weizmann Institute of Science, 76100 Rehovot, Israel

relatively large channels ($\simeq 0.5$ cm) of constant rectangular cross-section. Vinen's interesting observations, however, cannot give direct spatial information on generation of VLD, as his second sound system detected the whole volume of the channel simultaneously. Mendelssohn and Steele [2] were the first who claimed to observe of turbulent fronts filling a long thin circular thermal counterflow pipe, propagating into it from both the cold and hot ends. Many investigations followed; let us mention here detailed investigations of propagating and stationary turbulent fronts in rather narrow uniform as well as non-uniform counterflow channels performed by Tough's group (see [3] and references therein). These studies in narrow channels were based primarily on measurements of the temperature difference $\Delta T$ (and in some cases the chemical potential difference $\Delta \mu$ [4]) over the counterflow channel and therefore cannot provide direct information about spatiotemporal development of VLD in the channel.

In order to study the development and streamwise structure of the vortex tangle in counterflow QT, several localised probes are required. Such a setup is very difficult to realize experimentally in narrow channels of characteristic cross-section below $\approx 1$ mm$^2$, where two turbulent states, referred to as T1 and T2 have been observed - see, e. g., [4]. On the other hand, multiple turbulent states are generally not observed in channels of large cross-section of order $\approx 1$ cm$^2$. According to current understanding, the existence of multiple turbulent states is a consequence of the two fluid behaviour of superfluid $^4$He at temperatures above about 1 K, as either of its two components (or both of them, coupled by the mutual friction force) might become turbulent; and the outer scale of QT, i.e., the channel dimensions, ought to play a significant role. This paper report our first results of our planned complementary experimental, theoretical and numerical studies of counterflow QT, namely on the spatiotemporal development of VLD and its streamwise distribution in counterflowing $^4$He.

## 2 Experimental setup

We have designed and manufactured approximately 20 cm long, 7 mm-wide multipurpose channel of square cross-section, sketched in Fig. 1. The channel is equipped with three second-sound sensors (labelled L1, L2, and L3 hereafter) and seven additional ports, three of which in the present experiment are mounted with thermometers (labelled T1, T2, and T3; see Fig. 1). For second-sound sensors, standard design of capacitively actuated gold-plated Nuclepore membrane is used [5] and thermometry is accomplished using commercial Ge-on-GaAs film resistors. Using this system we are able to monitor initial local VLD build-up. It should be noted, however, that volumes probed by the second-sound transducers are macroscopic, approximately $7 \times 7 \times 7$ mm$^3$. Temperature data are more local, thermometers being just over 1 mm in extent.

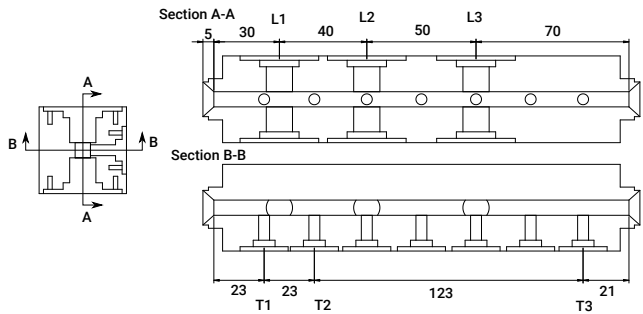

**Fig. 1** Sketch of the counterflow channel and the positions of the probes. The width of the flow cavity itself is 7 mm. Labels L$n$ indicate the positions of the second sound transducers and T$n$ the positions of the thermometers. The sockets on the channel, which do not have thermometers, are closed with blank holders geometrically identical to those with thermometers. The heater is mounted on the left side of the drawing. The drawing is to scale.

## 3 Experimental results

Typical growth of VLD is shown in Fig. 2. Initially a rapid transition for L1 and L2 is observed, followed by a slower, gradual increase to the eventual steady state, which is the only regime observed for L3. The initial transition can be either rapid increase (L2) or decrease (L1) of VLD. Negative vortex line density means simply that the attenuation of second sound decreased from its value in a quiescent helium before the flow was initiated, which is taken as a reference point. This probably corresponds to a partial removal of pinned remnant vortices, which is also supported by numerical simulations below. One might notice that there is a systematic relationship between steady state vortex line densities at the three sensor locations, namely VLD at L2 being highest, L1 being middle and L3 being lowest. This is most probably simply due to slightly differing quality of the sensors or minor discrepancies in the channel geometry and should not be taken as an evidence for streamwise inhomogeneity of the vortex tangle.

The standard procedure for obtaining time dependence of vortex line density is to excite the second sound resonator at one of its resonance frequencies and to monitor the real component of the complex amplitude of the oscillations (the imaginary being zero for an ideal resonator), for example by a lock-in amplifier. This requires that the resonance frequency does not change throughout the measurement. However, in the present experiment we observe temperature gradients which, along the fairly long channel, can cause appreciable temperature change (see Fig. 3 (a)) and cause the resonance peak to shift significantly (Fig. 3 (b)) thus rendering the standard procedure no longer applicable, except at 1.65 K where second sound velocity depends on temperature only very weakly. A common solution is to scan the amplitude over a small frequency range for every data point. However, this introduces temporal uncertainty on the points and decreases the data acquisition rate to a point,

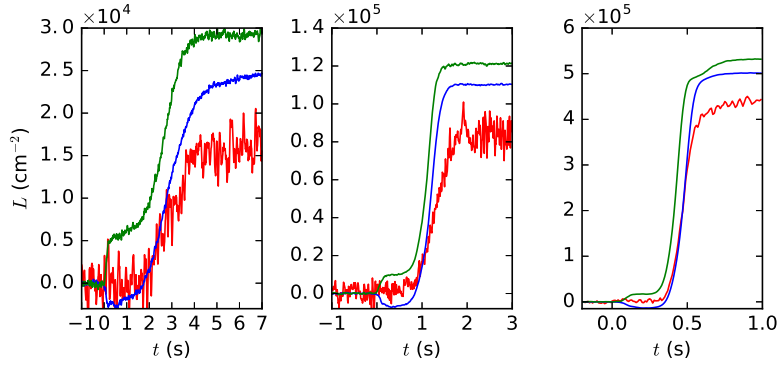

**Fig. 2** Typical growth of VLD at 1.65 K. VLD at position L1: blue (closest to the counterflow heater, middle line in steady state), L2: green (top in stead state), L3: red (bottom in steady state). All curves represent point-wise averages over 100 individual events. Savitzky-Golay smoothing has been applied to L3. Left, middle and right panels correspond to 20, 100, and 200 mW/cm$^2$ heat flux, respectively. (Color figure online)

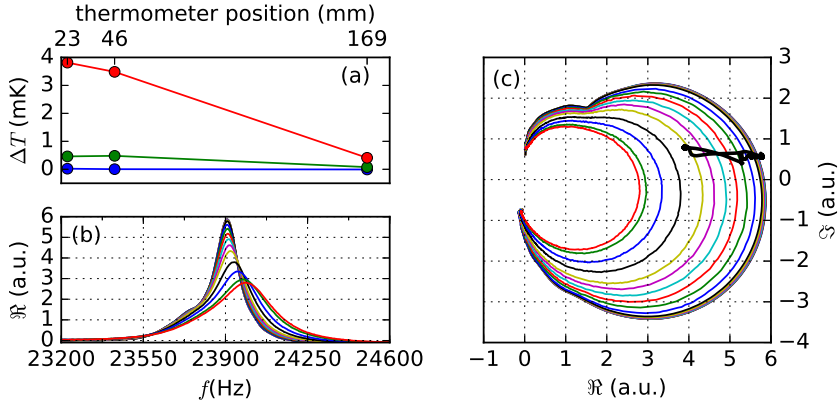

**Fig. 3** The temperature shift and its effect on the second sound measurement at 1.30 K. (a): the steady-state temperature shifts across the channel for approx. 20, 100, and 200 mW/cm$^2$ (lines from bottom to top). (b): Shift of the second-sound resonance peak. The curves span heat fluxes from 0 to about 300 mW/cm$^2$. (c): Determination of the resonance amplitude using off-resonance data. "Circles" are the same data as in (b) and the black superimposed curve is a typical experimental run – single case of buildup and decay.(Color figure online)

where it would be difficult, if not impossible, to resolve fast initial transients. Additionally, in our experience, this method is not applicable to fast ($< 0.5$ s) and large (around 50 Hz, with FWHM around 100 Hz) shifts. Therefore we propose a different approach, illustrated in Fig. 3 (c). The temperature shifts are sufficiently large to shift the peak, but small enough not to cause any other significant distortion. Thus the graphs of imaginary *versus* real component of the complex amplitude form a set of non-overlapping curves. Each of these curves has a resonance amplitude, thus it is possible, with suitable interpolation, to assign a well-defined amplitude to points in the complex plane spanned by the measured "calibration" resonance peaks, at least roughly within the resonance linewidth. In the tails of the resonance the sensitivity will necessarily decrease. This way, we can determine the resonance amplitude even when exciting the resonance slightly off-peak, without sacrificing temporal resolution.

## 4 Numerical Simulations

It is believed that QT is generated in the channel from "seed" vortices pinned to the channel walls or from some remnant vortices in the bulk of the channel. Schwarz [6], using a local induction approximation, showed that a wall-attached vortex near the channel inlet may serve as vortex mill to seed the flow with vortices. Indeed, it was very recently shown by large-scale Gross-Pitaevski simulations of the superfluid flow near realistic surface [7] that the critical velocity, necessary for nucleation of vortices, is easily exceeded near sharp excrescences of the surface.

**Method.** To understand different VLD build-up scenarios observed in the experiment, we perform numerical simulations using the Vortex Filament Method [8]. The simulations are set in a long plane channel of width $h = 0.1$ cm and length $H = 20h$, which is a scaled-down analog of the experimental channel. To generate the counterflow, we impose a prescribed parabolic time-independent profile of the streamwise projection of the normal velocity $V_z^n(y)$, oriented towards positive $z$-direction. The applied superfluid velocity is calculated dynamically at each time step to maintain the counterflow condition. Here we use the reconnection method [9] and the line resolution $\Delta\xi = 1.0 \times 10^{-3}$ cm. Open boundary conditions are used in the streamwise $z$ and periodic conditions in the spanwise $x$ directions. In the wall-normal $y$ direction, we impose $V_y^s(\pm h) = 0$ and $\boldsymbol{s}'(\pm h) = (0, \pm 1, 0)$ at the solid walls. The details of the simulation method may be found in Refs. [10,11]. To study the build-up of VLD and to compare it with the experimental results we calculate the streamwise VLD profile $\mathcal{L}(z)$, by accounting for the part of the lines configuration $\mathcal{C}'$ within a channel section of width $\Delta Z = 0.1$ cm, such that the channel is divided into 20 cubic sections of size $\mathcal{V}' = 0.1 \times 0.1 \times 0.1$ cm$^3$.

To quantify the degree of the anisotropy of the developing tangle, we use the profiles of the anisotropy indices[8]:

$$I_\parallel(z) = \frac{1}{\mathcal{L}(z)\mathcal{V}'} \int_{\mathcal{C}'} [1 - (\boldsymbol{s}' \cdot \hat{\boldsymbol{r}}_\parallel)^2] d\xi \,, \ I_\perp(z) = \frac{1}{\mathcal{L}(z)\mathcal{V}'} \int_{\mathcal{C}'} [1 - (\boldsymbol{s}' \cdot \hat{\boldsymbol{r}}_\perp)^2] d\xi \,, \quad (1)$$

where $\hat{\boldsymbol{r}}_\parallel$ and $\hat{\boldsymbol{r}}_\perp$ are unit vectors in the direction parallel and perpendicular to $\boldsymbol{V}^n$, respectively. In the isotropic case we have $I_\parallel = I_\perp = 2/3$. If all vortex lines lie in planes normal to $\boldsymbol{V}^n$, then $I_\parallel \approx 1, I_\perp \approx 1/2$, while in the steady state homogeneous vortex tangle under counterflow conditions $I_\parallel \approx 0.8, I_\perp \approx 0.6$ [8,10,12].

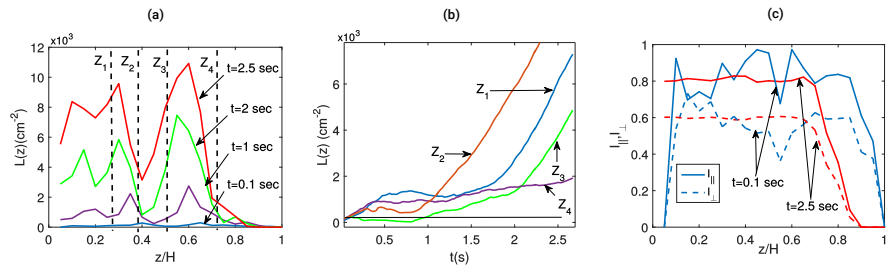

**Fig. 4** The averaged vortex lines density for realistic distribution of the seeding loops. Panel(a): the streamwise profiles at several time moments. Panel (b): time evolution at several locations. The horizontal black line marks the value of the mean initial density. Panel(c): The anisotropy indices profiles at several time moments. (Color figure online)

To mimic the experimental situation, we use initial conditions with seeding vortex loops placed on the walls at the positions corresponding to the sensors mounting points. Additionally, some loops are placed in the bulk, reflecting the fact that the critical velocity may be reached spontaneously [7] and these loops are moving inside the channel even in the absence of thermal counterflow. Such initial conditions produce a lightly inhomogeneous initial VLD distribution along the channel, ranging between 50 and 250 cm$^{-2}$. To make sure that the particular choice of the initial conditions does not influence the results, we use 20 such conditions that differ by exact position of the loops placement and their orientation. The resulting configurations are analyzed first individually and the resulting profiles further averaged. These simulations are carried out at $T = 1.65$ K and centerline normal velocity $V^n(0) = 1.5$ cm/s.

## 5 Numerical results

The averaged streamwise profiles of VLD along the channel at several time instants and the time evolution at several locations are shown in Fig. 4. The initial slight density inhomogeneity is enhanced (see panel(a)) and the initial stages of evolution are very different at different locations (panel(b)). In particular, at the position $Z_3$ the density is first diminished, as the nearby vortex-lines group is moving downstream, but then recovered as the tangle becomes dense and starts to grow. In other three locations the density growth includes some kind of a plateau, at a small value of $\mathcal{L}(z)$, with subsequent faster or slower growth. Different scenarios may be found at different locations. Notice that the evolution of the individual realizations are similar to the averaged, differing slightly in the exact locations of the density peaks and therefore, the positions where different growth patterns are realized.

The dynamics of the developing tangle is governed by competition between the downstream drift due to superfluid velocity and the diffusion of the vortices [13–15]. At $T = 1.65$ K, the superfluid velocity is sufficiently strong to cause overall downstream drift at initial stages of the evolution. Each group of vortices initially grows across the channel, forming thin sheets of vortex lines, moving downstream parallel to each other. In Fig. 4(c) we show the profiles of the anisotropy indices 1 at very early stages and at later time, when the tangle becomes fully three-dimensional. We see that initially the vortex lines are mostly oriented normally to the direction of the counterflow, while later the tangle becomes typically slightly oblate, but homogeneous along the entire channel.

# 6 Conclusions

We have experimentally probed the growth dynamics of the VLD at three locations inside a long counterflow channel of relatively large rectangular cross-section. Our preliminary data, obtained at 1.65 K, clearly show that the VLD builds up at different locations almost simultaneously and disproves the model according to which the turbulent front moves from any side of the channel with either superfluid or normal mean velocities, as it was indirectly observed and claimed in long but very thin channels. The complementary numerical simulations show that the experimentally observed complex patterns of the early VLD build-up is consistent with the tangle growth from multiple localized sources of remnant vortices. Initially, the vortex loops grow preferentially across the channel. At later stages, the tangle anisotropy becomes typical for the counterflow conditions and homogenous along the channel, despite inhomogeneous streamwise distribution of VLD.

**Acknowledgements** This work is supported by the Czech Science Foundation under project GAČR 203/14/02005S and by the European Community Framework Programme 7, EuHIT - European High-performance Infrastructures in Turbulence, grant agreement no. 312778.